\documentclass[11pt]{article}
\usepackage{amsmath,amssymb,color,slashed}
\usepackage[all]{xy}

\csname@addtoreset\endcsname{equation}{section}

\usepackage{amsfonts}

\newcommand{\hoch}[1]{$\, ^{#1}$}

\newcommand{\be}{\begin{equation}}
\newcommand{\ee}{\end{equation}}
\newcommand{\bea}{\setlength\arraycolsep{2pt} \begin{eqnarray}}
\newcommand{\eea}{\end{eqnarray}}
\newcommand{\nn}{\nonumber}

\def\ft#1#2{{\textstyle{\frac{\scriptstyle #1}{\scriptstyle #2} } }}
\def\fft#1#2{{\frac{#1}{#2}}}

\def\0{{\sst{(0)}}}
\def\1{{\sst{(1)}}}
\def\2{{\sst{(2)}}}
\def\3{{\sst{(3)}}}
\def\4{{\sst{(4)}}}
\def\5{{\sst{(5)}}}
\def\6{{\sst{(6)}}}
\def\7{{\sst{(7)}}}
\def\8{{\sst{(8)}}}
\def\sst#1{{\scriptscriptstyle #1}}

\def\ep{{\epsilon}}
\def\del{{\partial}}

\def\RR{{\mathfrak R}}
\def\MM{{\cal M}}

\def\cG{{{\cal G}}}
\def\Dslash{{\slashed D}}

\begin{document}

\begin{flushright}
\hfill{ \
MIFPA-11-27\ \ \ \ }
\end{flushright}

\vspace{25pt}
\begin{center}
{\large {\bf Critical and Non-Critical Einstein-Weyl Supergravity}}

\vspace{15pt}

H. L\"u\hoch{1,2}, C.N. Pope\hoch{3,4}, E. Sezgin$^3$ and L. Wulff$^3$

\vspace{10pt}

\hoch{1}{\it China Economics and Management Academy\\
Central University of Finance and Economics, Beijing 100081, China}

\vspace{10pt}

\hoch{2}{\it Institute for Advanced Study, Shenzhen
University\\ Nanhai Ave 3688, Shenzhen 518060, China}

\vspace{10pt}

\hoch{3} {\it George P. \& Cynthia Woods Mitchell  Institute
for Fundamental Physics and Astronomy,\\
Texas A\&M University, College Station, TX 77843, USA}

\vspace{10pt}
\hoch{4}{\it DAMTP, Centre for Mathematical Sciences,
 Cambridge University,\\  Wilberforce Road, Cambridge CB3 OWA, UK}

\vspace{40pt}

\underline{ABSTRACT}
\end{center}

   We construct ${\cal N}=1$ supersymmetrisations of some
recently-proposed theories of critical gravity, conformal gravity, and
extensions of critical gravity in four dimensions.
The total action consists of the sum of three separately off-shell 
supersymmetric actions containing Einstein gravity, a cosmological 
term and the square of the Weyl tensor.
For generic choices of the coefficients for these terms,
the excitations of the resulting theory around an AdS$_4$ background
describe massive spin-2 and massless spin-2 modes coming
from the metric; massive spin-1 modes coming from a vector field in the
theory; and massless and massive spin-$\ft32$ modes (with two unequal masses)
coming from the gravitino. These assemble into a massless and a massive
${\cal N}=1$ spin-2 multiplet. In critical supergravity, the coefficients
are tuned so that the spin-2 mode in the massive multiplet becomes massless.
In the supersymmetrised extensions of critical gravity, the coefficients
are chosen so that the massive modes lie in a ``window" of lowest energies
$E_0$ such that these ghostlike fields can be truncated by imposing
appropriate boundary conditions at infinity, thus leaving just positive-norm
massless supergravity modes.

\vspace{15pt}

\thispagestyle{empty}


\tableofcontents


\newpage

\section{Introduction}

   It was shown in \cite{stelle1,stelle2} that four-dimensional Einstein
gravity with additional curvature-squared terms is perturbatively
renormalisable.  The renormalisability comes at a price, namely that the
excitations around a Minkowski background contain states of negative norm
as well as states of positive norm.  Specifically, the excitations
comprise massive spin-0 and massless spin-2 modes with positive norm,
and massive spin-2 modes with negative norm.  By choosing the
curvature-squared terms to be of the form of the square of the Weyl tensor,
the spin-0 modes can be eliminated.  It was recently observed
that if a cosmological constant is added as well, the coefficient of
Weyl-squared can be adjusted so that
the massive spin-2 modes become massless \cite{lupo}.  This theory of
``critical'' gravity thus describes regular massless spin-2 excitations
and logarithmic spin-2 excitations around an AdS$_4$ background. The
energies of the massless spin-2 modes are zero, whilst those of the
logarithmic modes are in general nonvanishing \cite{lupo}.  However, as
discussed in \cite{porrob,lilulu}, the energies of the general excitations
can have either sign, and so one would have to truncate out the logarithmic
modes in order to avoid ghostlike modes.  This, unfortunately, leaves a
rather empty theory with only zero-norm massless spin-2 states.

    Maldacena recently considered the conformally-invariant theory
with a pure Weyl-squared action, in which
the massive spin-2 field in an AdS$_4$ background
is truncated by imposing an appropriate
boundary condition \cite{maldaconf}.  This is possible
because the massive spin-2 mode actually has a negative mass-squared in this
case, meaning
that it carries a non-unitary representation of $SO(2,3)$, but it is
not sufficiently negative to imply that it is tachyonic.  This massive
mode has a slower fall-off than the massless spin-2 mode, and so it can
be eliminated, while retaining the massless mode, by imposing a suitable
AdS fall-off condition at infinity.

   It was subsequently observed in \cite{lupapo} that there exists a
natural generalisation of critical gravity, in which the coefficient of
Weyl-squared that is added to cosmological Einstein gravity is chosen to
lie anywhere
in the range where the massive spin-2 mode has negative, but not tachyonic,
mass-squared.  This gives a one-parameter family of theories where one
can truncate out the ghostlike massive spin-2 modes by the imposition of
boundary conditions, while retaining the (positive norm) massless spin-2
modes.  One end of the parameter range corresponds to the pure Weyl-squared
theory considered by Maldacena.

   In this paper, we study an ${\cal N}=1$ supersymmetric extension of
cosmological gravity with the Weyl-squared term. We do this by starting
from known results for an off-shell chiral superfield formulation, and then
re-expressing the Lagrangian in a component field expansion. We shall work with
off-shell $D=4, {\cal N}=1$ supergravity  with the so-called
old minimal set of
auxiliary fields
\cite{Stelle:1978ye,Ferrara:1978em}\footnote{For higher derivative
off-shell $D=4$, ${\cal N}=1$
supergravity in the new minimal formulation, see
\cite{Ferrara:1988pd, Cecotti:1987qe,deRoo:1990zm,LeDu:1997us}.}.
Prior to adding
in the Weyl-squared multiplet, the off-shell theory of Einstein supergravity
plus cosmological constant contains an
auxiliary vector field and an auxiliary complex scalar field.
After adding in the
Weyl-squared terms the auxiliary vector becomes dynamical, with
propagating massive spin-1 modes. However, the complex scalar remains 
non-dynamical.

   In section 2, we perform a component expansion of the chiral superfield
expressions for the ${\cal N}=1$ off-shell supersymmetric actions whose
bosonic sectors correspond to
Einstein gravity, the cosmological term, and the square of the Weyl tensor.  
These are exactly the ingredients in critical gravity and its extensions.
For simplicity, we restrict attention to those terms that will contribute
when computing the linearised fluctuation equations around the AdS$_4$
vacuum.  In section 3, we derive the relevant equations of motion, and
the linearised equations for the fluctuations.  These give a fourth-order
equation for spin-2 fluctuations, a third-order equation for spin-$\ft32$
fluctuations, and a second-order equation for the spin-1 fluctuations.
In section 4, we analyse the multiplet structure for the fluctuation fields,
showing how, in general, they comprise a massless ${\cal N}=1$ spin-2
multiplet, and a massive ${\cal N}=1$ spin-2 multiplet.  We also analyse
the action of the supersymmetry transformations on the various fields.

  In section 5, we examine possible ways to obtain ghost-free theories.  This
can be achieved by choosing the coefficient of the Weyl-squared term so
that the undesirable negative-norm massive fields can be truncated from
the spectrum by the imposition of appropriate boundary conditions, while
still retaining the fields in the massless multiplet.  We consider two
cases; critical supergravity, where the massive multiplet becomes massless,
giving rise to logarithmic modes that can be truncated from the spectrum; and
a 1-parameter family of non-critical theories where the massive spin-2 fields
are all in non-unitary representations of the AdS algebra, and which therefore
have slower fall-off than the massless modes, allowing them again to be
truncated by a suitable boundary condition.  The paper ends with
conclusions in section 6.  In a set of three appendices we present some
of our notation and conventions; a detailed discussion of the ${\cal N}=1$
superspace constraints; and an explicit construction of the transformations,
using Killing spinors, that relate the spinor and tensor harmonics in AdS$_4$
for all spins $s\le 2$.

\section{Off-Shell Supersymmetrisation of Cosmological Einstein plus
Weyl-squared Gravity}

   There is a standard recipe for constructing a supersymmetric action
from any chiral superfield $r$. The Lagrangian is given by \cite{Binetruy:2000zx}
\begin{equation}
e^{-1}\mathcal L=\left(\ft12 D^\alpha
D_\alpha+i(\bar\psi_a\sigma^a)^\alpha
D_\alpha+\bar \MM+\bar\psi_a\bar\sigma^{ab}\bar\psi_b\right)r|+\mathrm{h.c.}\,,
\label{superf}
\end{equation}
where the notation $X|$ denotes the lowest component ($\theta$ independent)
in the $\theta$ expansion of the superfield $X$.
The standard supergravity action is obtained by taking $r=-3\RR$,
where $\RR$ is a chiral
superfield whose lowest component is $\RR| =\ft16 \MM$, where $\MM$ is a
complex scalar auxiliary field (see Appendix B).\footnote{We use $\RR$ rather
than the conventional $R$ to denote the superfield, to avoid confusion with
the Ricci scalar.}  The resulting Lagrangian
is \cite{Stelle:1978ye,Ferrara:1978em}
\begin{eqnarray}
e^{-1}\mathcal L_1=
\ft12 R
+\ft13(A^\mu A_\mu - \bar\MM \MM )+
\ft12 \bar\psi_\mu \gamma^{\mu\nu\rho} \psi_{\nu\rho}\,,\label{einstlag}
\end{eqnarray}
where $D_\mu$ is the Lorentz-covariant derivative,  $A_\mu$ is a real
auxiliary vector field that also comes from $\RR$, and
\be
\psi_{\mu\nu} = 2 D_{[\mu}\psi_{\nu]}\ .
\ee
(See appendices A and B for further notation and conventions.)
The Ricci scalar $R$ in (\ref{einstlag}) is constructed from a spin-connection
with added quadratic fermion torsion.  These additional terms will not
concern us here, since they will not contribute to the linearised equations
in an AdS$_4$ background.

Taking instead $r=1$,  equation (\ref{superf}) gives
\begin{eqnarray}
e^{-1}\mathcal L_2&=&\MM+\bar \MM -\bar\psi_\mu\gamma^{\mu\nu} \psi_\nu\,.
\label{cosmo}
\end{eqnarray}
In backgrounds where $\MM$ is constant, this is the supersymmetrisation of a
cosmological constant term.

   Finally the Weyl-squared invariant is obtained by
taking $r=-\frac{1}{4}W^{\alpha\beta\gamma}W_{\alpha\beta\gamma}$,
where $W_{\alpha\beta\gamma}$ is a chiral superfield whose lowest
component is proportional to the gravitino curvature (see Appendix B):
\bea
e^{-1}\mathcal L_3&=& C^{\mu\nu\rho\sigma} C_{\mu\nu\rho\sigma} -
\ft23 F^{\mu\nu} F_{\mu\nu}
-\ft43 \bar\psi^{\mu\nu}\Dslash \psi_{\mu\nu} +
 \ft43\bar\psi_{\mu\lambda} \gamma^{\mu\nu\rho} D_\rho
\psi_\nu{}^\lambda + \cdots \,,
\eea
where
\be
F_{\mu\nu} = 2\del_{[\mu} A_{\nu]}\ ,
\ee
and the ellipses denote
terms of the form $\psi^2 \times\nabla$(bosonic fields) and quartic fermion terms. (These
terms will vanish when we study the excitations around an AdS$_4$ background,
and so we shall not need to consider them in this paper.)
Note that the square of the Weyl tensor can be written in terms of the
Riemann and Ricci curvature as
\be
C^{\mu\nu\rho\sigma} C_{\mu\nu\rho\sigma} =
R^{\mu\nu\rho\sigma} R_{\mu\nu\rho\sigma} -2 R^{\mu\nu} R_{\mu\nu} +
\ft13 R^2\,.
\ee

There exists just one more independent curvature-squared invariant, 
modulo total derivatives, for which
the superfield $r$ is given by
\be
r= \left(\bar D^{\dot\alpha} \bar D_{\dot\alpha} -4\RR\right) \RR \RR^\dagger\ .
\ee
In components, this gives rise to an $R^2$ term as well as a kinetic term
for the real part of the auxiliary field $\MM$.  We shall
not consider this invariant further, in this paper, so that the scalar 
fields remain non-dynamical.

The off-shell supersymmetry transformation rules are
\bea
\delta e_\mu{}^a &=& \bar\epsilon \gamma^a\psi_\mu\,,\nn\\
\delta\psi_\mu &=& - D_\mu\epsilon - \frac{i}{6} (2 A_\mu +
 \gamma_{\rho\mu} A^\rho)\gamma_5 \epsilon -
 \ft16 \gamma_\mu(S+ i\gamma_5 P)\epsilon\,,\nn\\
\delta S &=& \bar\epsilon \gamma^{\mu\nu} \widehat\psi_{\mu\nu}\,,\nn\\
\delta P &=& i \bar\epsilon \gamma^{\mu\nu} \gamma_5\widehat\psi_{\mu\nu}\,,\nn\\
\delta A_\mu &=& \ft{i}8\bar\epsilon(\gamma_\mu
  \gamma^{\nu\rho} -3\gamma^{\nu\rho}\gamma_\mu)\gamma_5 \widehat\psi_{\nu\rho}\,,\label{susy1}
\eea
where
\be
\widehat\psi_{\mu\nu} = \psi_{\mu\nu}
+ \frac{i}{3} \gamma_5 \left(2 A_{[\mu} + A^\rho
 \gamma_{\rho[\mu} \right) \psi_{\nu]} +
 \ft13 \gamma_{[\mu}(S+ i\gamma_5 P)\psi_{\nu]}\,,
\ee
and
$\MM$ is written in terms of real scalar and pseudoscalar fields
as $\MM=S+i P$.

   Here we shall consider a linear combination of the supersymmetric
Lagrangians discussed above,
\be
{\cal L}= {\cal L}_1 + a\, {\cal L}_2 + b\, {\cal L}_3\,.
\ee
Thus the total bosonic Lagrangian is
\be
{\cal L}_B= \ft12 R + \ft13(A^\mu A_\mu - S^2-P^2) +2 a S +
  b\, C^{\mu\nu\rho\sigma} C_{\mu\nu\rho\sigma}
  - \frac{2b}3 F^{\mu\nu} F_{\mu\nu}\,,
\label{Lbos}
\ee
and the total fermionic Lagrangian (modulo terms that will vanish in the
AdS$_4$ background we shall consider) is
\be
{\cal L}_F = \ft12 \bar\psi_\mu \gamma^{\mu\nu\rho}
\psi_{\nu\rho}  -a\, \bar\psi_\mu\gamma^{\mu\nu} \psi_\nu
-\frac{4b}{3}\,  \bar\psi^{\mu\nu}\Dslash \psi_{\mu\nu} +
 \frac{4b}3 \,\bar\psi_{\mu\lambda} \gamma^{\mu\nu\rho} D_\rho
\psi_\nu{}^\lambda\,.\label{Lferm}
\ee

\section{Equations of Motion and Linearisation around AdS$_4$}

\subsection{Bosonic fields}

    The bosonic equations of motion, following from (\ref{Lbos}), are
\bea
&&S = 3a\,,\qquad P=0\,,\label{SPeq}\\
&&\nabla^\mu F_{\mu\nu} + \fft1{4b}\, A_\nu =0\,,\label{Feq}\\
 &&R_{\mu\nu} -\ft12 R g_{\mu\nu} + \ft13 (S^2+P^2- 6 a S) g_{\mu\nu}
+ \ft23(A_\mu A_\nu - \ft12 A^\rho A_\rho \nn g_{\mu\nu}) \nn\\
&&-\fft{8b}{3} (F_\mu{}^\rho\, F_{\nu\rho} -
   \ft14 F^{\rho\sigma} F_{\rho\sigma} g_{\mu\nu} ) +E_{\mu\nu}=0\,,
\label{geq}
\eea
where $E_{\mu\nu}$, the contribution to the Einstein equation
from the Weyl-squared term, is given by
\bea
E_{\mu\nu} &=& 8b(R_{\mu\rho}\, R_\nu{}^\rho -\ft14 R^{\rho\sigma}
  R_{\rho\sigma}\, g_{\mu\nu})
-\fft{4b}{3} \left[R\, (R_{\mu\nu} -\ft14 R\, g_{\mu\nu}) +
g_{\mu\nu}\, \square R -\nabla_\mu\nabla_\nu R\right]\nn\\
&&+ 4 b \left[ \square R_{\mu\nu} + \ft12 \square R\,
  g_{\mu\nu} - 2\nabla_\rho \nabla_{(\mu} R_{\nu)}{}^\rho\right] \,.
\label{Edef}
\eea
The maximally-symmetric vacuum solution of the bosonic equations of
motion is given by setting $A_\mu=0$, and taking $g_{\mu\nu}$ to be
the metric on AdS$_4$, satisfying
\be
R_{\mu\nu\rho\sigma}= -a^2(g_{\mu\rho} g_{\nu\sigma}-
          g_{\mu\sigma} g_{\nu\rho})\,,\qquad
R_{\mu\nu}=-3a^2 g_{\mu\nu}\,,\qquad R= -12 a^2\,.\label{AdS}
\ee

     We may then consider the equations for linearised bosonic fluctuations
around this background.  For the metric, we consider
$\delta g_{\mu\nu}=h_{\mu\nu}$, and define\footnote{All covariant 
derivatives in the expressions expanded around AdS$_4$ are understood 
to be covariant with respect to the AdS$_4$ background connection.}
\bea
\cG_{\mu\nu}^L &=& R^L_{\mu\nu} -\ft12 R^L\, g_{\mu\nu} + 3 a^2\,
h_{\mu\nu}
\,,\label{GL}\\
R^L_{\mu\nu} &=& \nabla^\lambda\nabla_{(\mu} h_{\nu)\,\lambda}
   -\ft12\square h_{\mu\nu} -\ft12 \nabla_\mu\nabla_\nu h\,,\label{RicL}\\
R^L&=& \nabla^\mu\nabla^\nu h_{\mu\nu} -\square h + 3a^2 h\,,\label{RL}
\eea
where $h\equiv g^{\mu\nu} h_{\mu\nu}$.
The linearised equation for $h_{\mu\nu}$ is then given by \cite{lupo}
\bea
(4b\, \square + 1+16a^2 b)\cG^L_{\mu\nu} -\fft{4b}{3}
(\nabla_\mu\nabla_\nu - g_{\mu\nu}\square - 3 a^2 g_{\mu\nu}) R^L=0\,.
\label{hmunulin}
\eea
Noting that $g^{\mu\nu} \cG^L_{\mu\nu}=-R^L$, we find that the
trace of (\ref{hmunulin}) gives simply
\be
R^L=0\,.\label{hlin}
\ee

     We may consider a 1-parameter family of possible gauge choices for
$h_{\mu\nu}$, of the form
\be
\nabla^\mu h_{\mu\nu}= c \nabla_\nu h\,,\label{cgauge}
\ee
where $c$ is a constant.
(de Donder gauge corresponds to $c=\ft12$.)  The trace equation (\ref{hlin})
then implies
\be
(c-1) \square h + 3 a^2 h=0\,.
\ee
If we choose $c=1$ in the gauge condition (\ref{cgauge}) then we immediately
deduce that $h=0$, as in \cite{lupo}.  If we instead take $c\ne1$, then
we can make residual coordinate transformations $\delta x^\mu=\xi^\mu$ with
$\xi_\mu=\del_\mu\xi$, which will therefore preserve the gauge condition
(\ref{cgauge}) provided that $\xi$ satisfies
\be
(c-1) \square \xi + 3 a^2 \xi=0\,.
\ee
Since the transformation of $h$ is given by $h\rightarrow h+ 2 \square\xi$,
and since $h$ and $\xi$ satisfy the same equation,
it follows that $\xi$ can be used in order to set $h$ to zero.  Thus
for any value of $c$, whether equal to 1 or not, the trace mode $h$ can
be eliminated by the gauge choice.  We shall assume from now on that
this has been done, and so $h_{\mu\nu}$ is in transverse traceless gauge,
\be
\nabla^\mu h_{\mu\nu}=0\,,\qquad h=0\,.
\ee
The full linearised equation (\ref{hmunulin}) for $h_{\mu\nu}$ then becomes
\cite{lupo}
\be
(\square + 2 a^2)\left(\square+ 4 a^2 + \fft1{4b}\right) h_{\mu\nu}=0\,.\label{spin2}
\ee

   Provided that the constant terms in the two factors
are unequal, the general solution to the fourth-order equation (\ref{spin2})
is just a linear combination of solutions to the two second-order equations.
To see this, suppose we have
$(\square+\lambda_1)(\square+\lambda_2) h_{\mu\nu}=0$.  This can
be written as
\be
(\square+\lambda_1) h^{(1)}_{\mu\nu}=0\,,\quad \hbox{where}\quad
(\square+\lambda_2) h_{\mu\nu}= h^{(1)}_{\mu\nu}\,.
\ee
Defining
\be
h_{\mu\nu}= h^{(2)}_{\mu\nu} +\fft1{\lambda_2-\lambda_1}\,
h^{(1)}_{\mu\nu}\,,
\ee
we see that provided $\lambda_2\ne\lambda_1$, the general solution to the
fourth-order equation is a linear combination of $h^{(1)}_{\mu\nu}$ and
$h^{(2)}_{\mu\nu}$ satisfying
\be
(\square+\lambda_1) h^{(1)}_{\mu\nu}=0\,,\qquad
(\square+\lambda_2) h^{(2)}_{\mu\nu}=0\,.
\ee
Thus, equation (\ref{spin2}) implies that generically
there are massless spin-2 modes satisfying
\be
(\square + 2 a^2) h_{\mu\nu}=0\,,\label{spin2a}
\ee
and additional massive spin-2 modes satisfying
\be
(\square+ 4 a^2 + \fft1{4b}) h_{\mu\nu}=0\,.\label{spin2b}
\ee
The degenerate case where $2a^2= 4a^2 +1/(4b)$, i.e. $b=-1/(8a^2)$,
which in fact corresponds
to critical gravity, will be discussed in detail later, in section 5.

   For the vector $A_\mu$, which vanishes in the AdS$_4$ background, the
fluctuation equation is just given by the Proca equation (\ref{Feq}).  Taking the
divergence, one therefore finds
\be
\nabla^\mu A_\mu=0\,,\qquad (\square +  3a^2 + \fft1{4b})
  A_\mu=0\,.\label{Aeq2}
\ee

\subsection{The gravitino equation}

   The gravitino equation of motion in the AdS$_4$ background follows from
(\ref{Lferm}):
\be
\gamma^{\mu\nu\rho}\psi_{\nu\rho}
-2a\gamma^{\mu\nu}\psi_\nu
-\frac{8b}{3}\Big[
2\gamma^\rho D_\nu D_\rho \psi^{\mu\nu}
+\gamma_\nu{}^{\rho\sigma}D_\rho D_\sigma\psi^{\mu\nu}
-\gamma^{\mu\rho}{}_\nu  D^\sigma D^\nu \psi_{\rho\sigma}
\Big] =0\,.\label{fermioneq}
\ee
Multiplying with $\gamma_\mu$, and using the identity
$D_{[\mu}\psi_{\nu\rho]}=-\ft12 a^2 \gamma_{[\mu\nu} \psi_{\rho]}$ in the
AdS$_4$ background, we obtain
\be
D^\mu\psi_\mu - (\Dslash-\ft32 a)(\gamma^\mu\psi_\mu)=0\,.
\ee
Imposing the gauge condition $\gamma^\mu\psi_\mu=0$ implies also
$D^\mu\psi_\mu=0$, and the gravitino equation of motion (\ref{fermioneq})
gives
\be
\Dslash \square\psi_\mu + \left(3a^2+\fft1{4b}\right) \Dslash \psi_\mu +
   \fft{a}{4b}\psi_\mu=0\,.\label{feq2}
\ee
Using $(\Dslash)^2 \psi_\mu= \square\psi_\mu + 4a^2 \psi_\mu$, we can
rewrite (\ref{feq2}) in the factorised form
\be
(\Dslash + a)\left(\Dslash -\ft12 a-\ft12\sqrt{a^2- b^{-1}}\right)
 \left(\Dslash -\ft12 a +\ft12\sqrt{a^2-b^{-1}}\right)\psi_\mu=0\,.\label{ferm3}
\ee

   The analysis of this third-order equation is analogous to our earlier
discussion for spin 2.  Provided that the three constant terms in the
factorised form (\ref{ferm3}) are unequal, the general solution will be
a linear combination of the solutions to the three separate factors.  In
other words, there will be the massless gravitino mode
satisfying
\be
(\Dslash + a) \psi_\mu=0\,,\label{gravitinoa}
\ee
and two massive gravitino modes, satisfying, respectively,
\bea
&& (\Dslash -\ft12 a-\ft12\sqrt{a^2- b^{-1}}) \psi_\mu=0\,,\label{gravitinob}
\\
&&(\Dslash -\ft12 a+\ft12\sqrt{a^2- b^{-1}}) \psi_\mu=0\,,\label{gravitinoc}
\eea
The degenerate cases, where two eigenvalues coincide, will be treated
later in our discussion in section 5.

\subsection{The linearised supersymmetry transformations}

  We begin by observing that the AdS$_4$ background given by (\ref{AdS})
is supersymmetric.  This can be seen from the expression for
$\delta\psi_\mu$ in (\ref{susy1}), which vanishes in the AdS$_4$ background for
any Killing spinor solution $\epsilon_-$ of
\be
D_\mu \epsilon_\pm = \pm \ft12 a \gamma_\mu\epsilon_\pm\,.
\ee
In what follows, it will be understood when we use $\ep$ to denote a Killing
spinor, that it will be of the $\ep_-$ type.

   The linearised transformation rules, which will be useful for describing
how supersymmetry acts on the fluctuation modes, are given by
\bea
\delta h_{\mu\nu} &=& 2\bar\epsilon \gamma_{(\mu}\psi_{\nu)}\,,\nn\\
\delta\psi_\mu &=& \ft14 \nabla_\rho h_{\mu\sigma}\, \gamma^{\rho\sigma}
\epsilon -\fft{i}{6} (2 A_\mu + \gamma_{\rho\mu} A^\rho)\gamma_5\epsilon
-\ft14 a  h_{\mu\nu}\gamma^\nu \epsilon\,,\nn\\
\delta A_\mu &=& \ft{3}2 i \bar\epsilon\gamma_5 (\Dslash + a)\psi_\mu\,.
\label{linsusy}
\eea
In obtaining the expression for $\delta A_\mu$, we have used the
gauge condition $\gamma^\mu\psi_\mu=0$, and its consequence that
$D^\mu\psi_\mu=0$.

\section{Spectrum and Multiplet Structure of the Fluctuations}

   In this section, we investigate the structure of the
small fluctuations around the AdS$_4$ background, showing how the various
modes assemble into ${\cal N}=1$ multiplets under AdS supersymmetry.

\subsection{AdS representations of the fluctuations}

    Subject to appropriate boundary conditions, the solutions of the
linearised equations obtained in the previous section form unitary
irreducible representations of the $SO(3,2)$ AdS group.  These
representations, denoted by $D(E_0,s)$,  are labelled by their
lowest energy $E_0$ and their spin $s$.  The unitary irreducible
representations of ${\cal N}=1$ AdS supersymmetry fall into four
disjoint classes \cite{heid}, namely
\bea
\hbox{Class 1}:\quad &&D(\ft12,0) \oplus D(1,\ft12)\,,\nn\\
\hbox{Class 2}:\quad && D(E_0,0)\oplus D(E_0+\ft12, \ft12)\oplus
    D(E_0+1,0)\,,\qquad E_0>\ft12\,,\nn\\
\hbox{Class 3}:\quad && D(s+1,s) \oplus D(s+\ft32, s+\ft12)\,,\qquad
 s=\ft12,1,\ft32,\ldots \,,\\
\hbox{Class 4}:\quad && D(E_0,s)\oplus D(E_0 \! +\! \ft12,s \! +\!\ft12)\oplus
    D(E_0 \! +\! \ft12,s \! -\! \ft12)\oplus D(E_0\! +\! 1,s)\,,\quad E_0>s+1
\,.\nn
\eea
Class 1 is the singleton, supermultiplet; Class 2 is the Wess-Zumino
supermultiplet; Class 3 comprises massless gauge supermultiplets; and Class 4
comprises massive supermultiplets.

   The representations arising in our case can
be determined from the eigenvalues of the D'Alembertian (for bosons)
or the Dirac operator (for fermions).  For the fields of spins 2, 1 and
$\ft32$ of interest to us, one has
\bea
D(E_0,2):\qquad\qquad
    && \square h_{\mu\nu} = a^2 [E_0(E_0-3) -2] h_{\mu\nu}\,,\nn\\
D(E_0,1):\qquad\qquad
   && \square A_\mu = a^2 [E_0(E_0-3) -1] A_\mu\,,\\
D(E_0,\ft32):\qquad\qquad
   && \Dslash \psi_\mu^{\pm}= \pm a(E_0-\ft32)\psi_\mu^{\mp}\ ,\nn
\eea
where $\psi_\mu^{\pm}=\frac12(1\pm\gamma_5)\psi_\mu$.

    Let us first consider the general situation, for generic values of
the coefficient $b$ associated with the Weyl-squared term.  From
(\ref{spin2}) we see that there are
always massless spin-2 modes satisfying (\ref{spin2a}), in the
$D(3,2)$ representation, and from (\ref{ferm3}) there are always
massless spin-$\ft32$ modes satisfying (\ref{gravitinoa}), in the
$(\ft52,\ft32)$ representation.  These bosonic and fermionic modes
form the massless supermultiplet
\be
D(\ft52,\ft32) \oplus D(3,2)\,,
\ee
which is of Class 3 with $s=\ft32$.

   The remaining modes that we read off
from (\ref{spin2b}) for spin-2, (\ref{Aeq2}) for spin-1, and
(\ref{gravitinob}) and (\ref{gravitinoc}) for
spin-$\ft32$, can then be seen, respectively, to have
the $E_0$ values
\bea
\hbox{Spin-2}:\qquad && E_0= \ft32 \pm \ft12 \sqrt{1-\fft{1}{a^2 b}}\,,\nn\\
\hbox{Spin-1}:\qquad && E_0= \ft32 \pm \ft12 \sqrt{1-\fft{1}{a^2 b}}\,,\nn\\
\hbox{Spin-$\ft32$}:\qquad && E_0=2 \pm \ft12 \sqrt{1-\fft{1}{a^2 b}}\,,\quad
\hbox{and}\quad E_0=1\pm\ft12 \sqrt{1-\fft{1}{a^2 b}}\,.\label{massive}
\eea
When the plus sign is chosen in front of all the square roots, and if the
parameter $b$ is chosen so that
\be
\sqrt{1-\fft{1}{a^2 b}}  >3\,,\label{Ebound}
\ee
i.e so that
\be
-\fft1{8a^2} < b<0\,,
\ee
then the representations in (\ref{massive}) all satisfy the bound $E_0>s+1$,
and they can be seen to
form an ${\cal N}=1$ unitary massive supermultiplet, of the Class 4 type.
If (\ref{Ebound}) is not satisfied, then the multiplet will be non-unitary.
There is another massive multiplet, which is always non-unitary,
corresponding to taking
the minus sign in front of all the square roots.

   If the parameter $b$ lies in the range where $1-1/(a^2 b)$ is negative,
then the $E_0$ values become complex.  Since, in particular, the modes have
time dependence proportional to $e^{i E_0 t}$, this would imply that they
would have real exponential growth, corresponding to classical instability.
Such modes are tachyonic, and are the higher-spin analogues of scalar
modes that violate the Breitenlohner-Freedman bound \cite{breifree}.  We
shall always require that $b$ be chosen so that
\be
1-\fft1{a^2 b}\ge 0\,.\label{tachyon}
\ee

\subsection{Action of supersymmetry on the fluctuation modes }

   In this subsection we shall study the manner in which supersymmetry maps the
solutions of different spins into each other.  There are two reasons why
it is of interest to do this.  Firstly, it provides a simple way to obtain
explicit expressions for the solutions for all spins $s\le 2$, starting from
those for any particular given spin.  Secondly, it will give nontrivial
information about the multiplet structure, including
in the critical case, which we
shall discuss in section 5, when non-standard representations with
logarithmic behaviour arise. In the present section, we shall consider
just the non-critical case.

   We can determine how supersymmetry acts on the fluctuations by making
use of the linearised supersymmetry transformations given in equations
(\ref{linsusy}).  Essentially, we substitute a mode of one of the fields,
satisfying (\ref{spin2a}), (\ref{spin2b}), (\ref{Aeq2}), (\ref{gravitinoa}),
(\ref{gravitinob}) or (\ref{gravitinoc}), into the right-hand sides of
the transformation rules, and thus read off the associated
supersymmetry-related modes.  To be precise, it is necessary also to
make appropriate compensating gauge transformations (general coordinate,
and/or local Lorentz), in order to ensure that the supersymmetry-related
modes obey the appropriate gauge conditions we are imposing, which amount to
their being divergence-free and ($\gamma$--)traceless.

  To begin, we observe that if $\psi_\mu$ satisfies the massless gravitino
equation (\ref{gravitinoa}), then the $\delta h_{\mu\nu}$ transformation in
(\ref{linsusy}) generates a massless spin-2 solution, since
\be
(\square+2a^2) [2\bar\epsilon\gamma_{(\mu}\psi_{\nu)} + \delta_\xi h_{\mu\nu}]
  =0\,,
\ee
where the compensating general coordinate transformation is given by
\be
\delta_\xi h_{\mu\nu} = 2\nabla_{(\mu}\xi_{\nu)}\,,\qquad
 \xi_\mu= \fft1{3a}\, \bar\epsilon \psi_\mu\,.
\ee
Note that the massless $\psi_\mu$ mode does not generate any spin-1 solution,
since the $(\Dslash+a)$ operator in the $\delta A_\mu$
transformation in (\ref{linsusy}) annihilates the massless gravitino solution.

   In the reverse direction, substituting the massless spin-2 solution
(\ref{spin2a}) into the $\delta\psi_\mu$ transformation, we find that
indeed
\be
(\Dslash + a) [\ft14 \nabla_\rho h_{\mu\sigma}\, \gamma^{\rho\sigma} \ep
 - \ft14 a h_{\mu\nu}\, \gamma^\nu\ep]=0\ ,
\ee
which shows that the $\delta\psi_\mu$ generates a massless spin-3/2  solution.

   By similar reasoning, we find that the solutions of the massive fluctuation
equations map into one another under the linearised supersymmetry
transformations, forming the massive supermultiplet that we discussed in the
previous subsection. In this case, the required compensating general coordinate transformation
is given by
\be
\delta_\xi h_{\mu\nu} = 2\nabla_{(\mu}\xi_{\nu)}\,,\qquad
 \xi_\mu= \frac{1}{2a-\lambda} \, \bar\epsilon \psi_\mu\ ,
\label{compensating-h}
\ee
where $\Dslash \psi_\mu=\lambda\psi_\mu$ with $\lambda$ to be read off
from (\ref{gravitinob}) and (\ref{gravitinoc}). The singular situation
where $\lambda=2a$
arises in the critical case which will be discussed in
section \ref{critsec}.
The substitution of a massive gravitino solution into the right
hand side of $\delta A_\mu$
generates the solution for the massive vector field obeying the Proca
field equation (\ref{Aeq2}).
Finally, with the substitution of the massive graviton solution into
$\delta \psi_\mu$, it solves the equation
\be
(\Dslash -\ft12 a-\ft12\sqrt{a^2- b^{-1}})
 (\Dslash -\ft12 a +\ft12\sqrt{a^2-b^{-1}})
[\ft14 \nabla_\rho h_{\mu\sigma}\, \gamma^{\rho\sigma} \ep
 - \ft14 a h_{\mu\nu}\, \gamma^\nu\ep]=0\,,
\ee
and thus both of the massive gravitino modes arise, as a linear combination. Substituting the massive  spin-1 solution in $\delta\psi_\mu$ on the other hand, again yields a linear combination of  massive gravitino solutions, provided that we take into account compensating supersymmetry transformation needed to ensure that $\delta\psi_\mu$ is divergent-free and $\gamma$-traceless. This compensating transformation, whose parameter we shall denote by $\hat\epsilon$ is given by
\be
\hat\epsilon = \frac{ib}{3(1+8a^2b)} \left(4a A_\mu\gamma^\mu-F_{\mu\nu}\gamma^{\mu\nu}\right)\gamma^5 \epsilon\ .
\label{compensating-psi}
\ee
Note that only terms that are linear in fluctuation fields are to be retained in $\delta_{\hat\epsilon}\psi_\mu$.
Moreover, the overall factor is divergent at the critical point
that will be discussed further in section \ref{critsec}.

In summary, we have shown that away from the critical point the
fluctuations form a massless and a massive supergravity multiplet,
both on shell, as shown in the figure below, where the superscripts
refer to massive states whose AdS energies are given in (\ref{massive}).\\

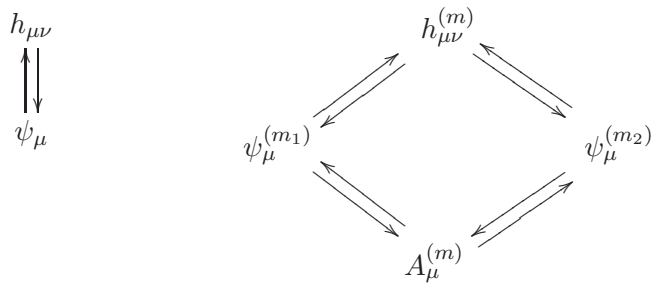
\begin{displaymath}
    \xymatrix{
    h_{\mu\nu}\ar@<.5ex>[d]  \\
      \psi_\mu\ar@<.5ex>[u]
        }\qquad\qquad\qquad
    \xymatrix{
                                  & {}\quad h_{\mu\nu}^{(m)} \ar@<1ex>[dl]\ar[dr] &   \\
        \psi_\mu^{(m_1)}\ar[dr]\ar[ur]   &                           &{}\quad \psi_\mu^{(m_2)}\ar[dl]\ar@<-1ex>[ul]\\
                                   & A^{(m)}_\mu\ar@<-1ex>[ur]\ar@<-1ex>[ul]               &  }
\end{displaymath}

\section{Ghost-free Supergravities}

   As is well known in the case of pure cosmological
gravity with a Weyl-squared term, the massive spin-2 excitations around the
AdS$_4$ background have energies that are opposite in sign to those of the
massless spin-2 modes (see, for example, \cite{lupo}).
Thus if the overall sign of the action is chosen
so that the massless graviton has positive-energy excitations, then the
massive spin-2 modes will be ghostlike.  In order to
achieve a ghost-free theory, one may try to
eliminate the massive excitations by imposing some appropriate boundary
conditions at infinity.  The situation for the supersymmetric extensions
that we are considering in this paper is similar, and so we can again
examine the circumstances under which such a truncation of the massive
multiplets may be possible.

   It is useful to divide the discussion into two cases.  One case arises
when the critical choice for the parameter $b$ is taken, namely
\be
b=b_{\rm crit} = -\fft1{8 a^2}\,.
\ee
In this case, the massive spin-2 modes, satisfying (\ref{spin2b}), become
massless, resulting in the emergence of a new type of solution to the
fourth-order equation (\ref{spin2}) that has a logarithmic dependence on the
radial AdS$_4$ coordinate.  An analogous phenomenon occurs also in the
spin-$\ft32$ sector. We shall discuss this case in subsection \ref{critsec}
below.  The logarithmic modes have indefinite norm, and must therefore be
truncated out in order to achieve a ghost-free theory. However, the massless
spin-2 modes have zero norm in this case \cite{lupo}, and so after the
truncation one is left with a rather trivial theory. A further feature, in
this critical case, is that the kinetic term $-\ft23 b F^{\mu\nu} F_{\mu\nu}$
for the spin-1 fields has the ``wrong sign.''

   The second case, which we shall consider first, corresponds to the
situation where $b$ is instead chosen so that the unitarity bound
(\ref{Ebound}) is violated, while still respecting the condition
(\ref{tachyon}) for avoiding tachyons.  This will provide a supersymmetric
generalisation of the ``extended critical gravities'' considered recently
in \cite{lupapo}.

\subsection{Extensions of critical supergravity}\label{noncritsec}

   In order to be able to impose boundary conditions that eliminate the
ghost-like massive modes, while retaining the desired massless modes, it
is necessary to choose the $b$ parameter to lie in a range where the massive
modes have a slower fall-off at infinity than the massless modes.
The fall-off is governed by the lowest-energy eigenvalue $E_0$, with modes
having larger $E_0$ falling off faster than those with smaller $E_0$. (See
for example \cite{Bergshoeff:2011ri}, where the spin-2 modes are constructed.)
Thus the desired choices for the parameter $b$ will be those for which the
massive modes are all non-unitary, satisfying $E_0<s+1$, while, by contrast,
the massless modes satisfy $E_0=s+1$.  Bearing in mind that we must still
require the massive modes to be non-tachyonic, in order to avoid classical
instabilities, it follows from (\ref{Ebound}) and (\ref{tachyon}) that
$b$ should be chosen to satisfy
\be
b \ge \fft1{a^2}\qquad \hbox{or}\qquad b\le -\fft{1}{8 a^2} \,.\label{noncrit}
\ee

   There is a further requirement, which excludes the negative $b$ choices in
(\ref{noncrit}).  This can be seen from the results in \cite{lupo,lupapo},
where the energies of the spin-2 modes are calculated.  In order to have
non-negative energies for the massless spin-2 modes, it is necessary that
$b$ should satisfy $b\ge -1/(8 a^2)$.  Thus we are led to consider the
1-parameter family of theories for which
\be
b \ge \fft1{a^2}\,.\label{noncritb}
\ee
For all values of $b$ within this range, the modes in the massive
supermultiplet will fall off more slowly than those in the massless
supermultiplet, and so they can be eliminated by imposing appropriate
boundary conditions at infinity.  Included in this family is the limit where
$b$ goes to infinity; after making an overall rescaling with a factor
$1/b$, this corresponds to the conformally-invariant case that is the
${\cal N}=1$ generalisation of the pure Weyl-squared gravity that was
recently considered by Maldacena \cite{maldaconf}.  In the entire range
(\ref{noncritb}), the excitations in the massless supermultiplet will all
have positive energies.

   It is interesting to note that at the lower end of the range in
(\ref{noncritb}), when $b=1/a^2$, the two massive spin-$\ft32$ branches in
(\ref{gravitinob}) and (\ref{gravitinoc}) become degenerate, and so
there will be spin-$\ft32$ modes with logarithmic coordinate dependence
in this case, even though none of the other members of the massive
supermultiplet will exhibit such behaviour.  It is also worth remarking that
the kinetic term $-\ft23 b F^{\mu\nu} F_{\mu\nu}$ for the spin-1 field
has the correct sign throughout the range (\ref{noncritb}).

\subsection{Critical supergravity}\label{critsec}

At the critical point we have
\begin{eqnarray}
b_{\rm crit}&=&-\frac{1}{8a^2}
\end{eqnarray}
and the linearized equations of motion become
\bea
\left(\square +2a^2\right)^2h_{\mu\nu} &=&0\ ,\\
\left(\square + a^2\right) A_\mu &=& 0\ ,\\
(\slashed D+a)^2(\slashed D-2a)\psi_\mu &=& 0\ .
\eea
It immediately follows that the vector field describes a massive
spin-1 mode with AdS energy $E_0=3$. As for the graviton and gravitino
field equations, to begin with they describe modes that follow from the
factorization of their wave operators. These are the massless spin-2 and
massless spin-3/2 modes, and a massive spin-3/2 mode
satisfying $(\slashed D -2a)\psi_\mu=0$, thereby having AdS
energy $E_0=7/2$. In addition to these, however, there will also
be logarithmic modes that satisfy the relations
\begin{eqnarray}
(\Box+2a^2)^2 h_{\mu\nu}^{\mathrm{log}}=0\,,&\qquad&(\Box+2a^2)h_{\mu\nu}^{\mathrm{log}}\neq0\,,\nonumber\\
(\slashed D+a)^2\psi_\mu^{\mathrm{log}}=0\,,&\qquad&(\slashed D+a)\psi_\mu^{\mathrm{log}}\neq0\ .
\label{log-modes}
\end{eqnarray}

Next we discuss how supersymmetry relates the fluctuation modes to each
other, to determine the underlying multiplet structure. As in our previous
discussion for the case of a generic massive multiplet,  we look at the
linearised supersymmetry transformations in (\ref{linsusy}), plug in the
various modes at the critical point on the right--hand--side and then
verify that the result satisfies an appropriate equation. In some cases
the supersymmetry transformations have to be accompanied by an appropriate
compensating gauge-transformation to preserve the gauge-conditions.

As we have seen previously, when we substitute the critical massive
gravitino mode satisfying $(\slashed D -2a)\psi_\mu=0$ into
$\delta h_{\mu\nu}$,
the compensating gauge-transformation (\ref{compensating-h}) that is needed
in order to preserve the gauge condition diverges. This means that
supersymmetry does not map the critical massive gravitino mode to a
transverse traceless 
spin-2 mode. Similarly we have seen that when the critical massive vector
mode is substituted in $\delta\psi_\mu$, the required compensating
gauge transformation (\ref{compensating-psi}) again diverges, which means
that the critical massive vector mode is not mapped to a gravitino mode in
the $\gamma^\mu\psi_\mu=0$, $D^\mu\psi_\mu=0$ gauge
by supersymmetry. When substituted into $\delta A_\mu$, the critical
massive gravitino will however give rise to a critical massive spin-1
mode, as follows immediately from the analysis we gave previously.

It remains only to analyse what happens when the logarithmic modes
satisfying (\ref{log-modes}) are substituted into the supersymmetry
transformations. Let us start with the supersymmetry variation of the
vector. It is not hard to verify that
\begin{equation}
(\Box+a^2)\left(\bar\epsilon\gamma_5(\Dslash + a)
\psi_\mu^{\mathrm{log}}\right)=0\,.
\end{equation}
This means that the gravitino log mode is mapped by supersymmetry into
the critical massive spin-1 mode. Next we consider what happens when the
graviton log mode is substituted into $\delta\psi_\mu$. Since the log modes
satisfy the same gauge conditions as the regular modes, no compensating
gauge transformation is needed and one finds
\be
(\slashed D+a)[\ft14 \nabla_\rho h_{\mu\sigma}^{\mathrm{log}}\, \gamma^{\rho\sigma} \ep
 - \ft14 a h_{\mu\nu}^{\mathrm{log}}\, \gamma^\nu\ep]=\ft14(\Box+2a^2)h_{\mu\nu}^{\mathrm{log}}\, \gamma^\nu\ep\neq0\ ,
\ee
and
\be
(\slashed D+a)^2(\slashed D-2a)[\ft14 \nabla_\rho h_{\mu\sigma}^{\mathrm{log}}\, \gamma^{\rho\sigma} \ep
 - \ft14 a h_{\mu\nu}^{\mathrm{log}}\, \gamma^\nu\ep]=0\,.
\ee
This shows that the graviton log mode is mapped by supersymmetry to a
linear combination of the gravitino log mode and the critical massive
gravitino mode. Finally we need to analyse what happens when the
gravitino log mode is substituted into $\delta h_{\mu\nu}$. In this case
a compensating general coordinate transformation will be needed to
preserve the gauge condition. With some work, one can show that
\be
(\square+2a^2)^2[2\bar\epsilon\gamma_{(\mu}\psi_{\nu)}^{\mathrm{log}} + \delta_\xi h_{\mu\nu}]=0\,,
\ee
where the compensating general coordinate transformation takes the form
\begin{equation}
\delta_\xi h_{\mu\nu} = 2\nabla_{(\mu}\xi_{\nu)}\,,\qquad
\xi_\mu=\frac{1}{9a^2}\bar\epsilon(\slashed D+4a)\psi_\mu^{\mathrm{log}}\,.
\end{equation}
This shows that the gravitino log mode is mapped by supersymmetry into
the graviton log mode. This completes the analysis of the supermultiplet
structure at the critical point. In addition to the massless supergravity
multiplet we have the non-standard multiplet
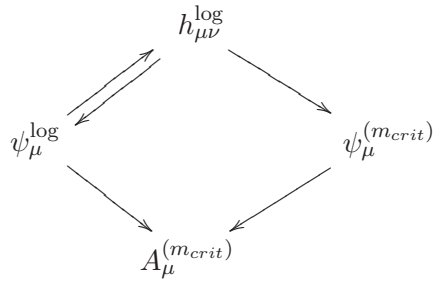
\begin{displaymath}
    \xymatrix{
                                  & {}\quad h_{\mu\nu}^{\mathrm{log}} \ar@<1ex>[dl]\ar[dr] &   \\
        \psi_\mu^{\mathrm{log}}\ar[dr]\ar[ur]   &                          &{}\quad \psi_\mu^{(m_{crit})}\ar[dl]\\
                                   & A_\mu^{(m_{crit})}              &  }
\end{displaymath}
where supersymmetry transformations are indicated by the arrows.

 Note that the logarithmic modes are not eigenstates of the AdS  energy
generator. Indeed, all of these modes are given as a product of a
universal logarithmic dependent factor and the solution for the
massless mode as \cite{Bergshoeff:2011ri}
\be
\phi^{\rm log} =  
  (2it + \log \sinh 2\rho -\log\tanh\rho ) \phi^{\rm massless}\ ,
\ee
where $\phi$ generically denotes any field that has
logarithmic mode, in a coordinate system in which the AdS$_4$ metric
is given by
\be
a^2 ds^2 = -\cosh^2\rho\, dt^2 + d\rho^2 +
\sinh^2\rho \left[ d\theta^2 + \sin^2\theta\, d\varphi^2 \right]\ .
\ee
We are not aware of a group theoretical analysis of the representations
of AdS superalgebra which accommodates such states. The analysis of
supersymmetry transformations, nonetheless, seems to suggest that that
if boundary conditions that exclude logarithmic modes are to be imposed,
then the full multiplet containing these modes are
to be excluded. In any event, in view of the recent developments in
the study of the critical bosonic gravity dynamical content
\cite{maldaconf,lupapo}, we shall not pursue further the supersymmetric
version of the story here.


\section{Conclusions}

   In this paper, we have constructed an ${\cal N}=1$ supersymmetrisation of
a class of four-dimensional gravities with a quadratic curvature modification
proportional to the square of the Weyl tensor.  The resulting supergravities
encompass supersymmetrisations of critical gravity \cite{lupo}, where the
coefficient of Weyl-squared is adjusted so that the generically massive
spin-2 excitations become massless; pure conformally-invariant Weyl-squared
gravity, which was recently proposed in \cite{maldaconf} as providing
an equivalent description of ordinary gravity in the long-wavelength regime;
and a class of generalisations of critical gravity considered recently in
\cite{lupapo}.

   We showed that the excitations of the ${\cal N}=1$ theory around its
AdS$_4$ vacuum generically describe a massless spin-2 multiplet and a
massive spin-2 multiplet.  In the critical gravity limit, the massive spin-2
field becomes massless, leading to the emergence of spin-2 and spin-$\ft32$
modes with logarithmic coordinate dependence.  The formerly massive
multiplet becomes a non-standard one in this limit, which lies outside the
usual classification of unitary ${\cal N}=1$ representations described in
\cite{heid}.

   The extensions beyond the critical limit, which are the supersymmetric
generalisation of the theories considered in \cite{lupapo}, arise when
the coefficient of the Weyl-squared term is chosen to lie in the range
where the massive fields carry non-unitary representations of $SO(2,3)$.  For
the bosons (spin-2 and spin-1), this means that they have mass-squared
values, defined as $(\Box + 2 a^2 -M_2^2) h_{\mu\nu}=0$ and
$(\Box + 3 a^2 -M_1^2) A_\mu=0$, that are negative. They are, however, not
sufficiently negative to be tachyonic, meaning that their lowest energies
$E_0$, given by
\be
E_0^{(2)} = \fft32 \pm \sqrt{\fft94 + \fft{M_2^2}{a^2}}\,,\qquad
E_0^{(1)} = \fft32 \pm \sqrt{\fft14 + \fft{M_1^2}{a^2}}\,,
\ee
are still real.  Because the lowest energies of the massive fields
all violate the unitarity bounds $E_0^{(s)} \ge s+1$, they have a slower
fall-off at large distance than the massless fields, and thus they
can be eliminated, while retaining the massless fields, by imposing
appropriate boundary conditions.  The same is true also for the logarithmic
modes in the case of critical gravity.  Eliminating the massive or
logarithmic modes is desirable from a physical point of view, since they
can have negative norms, and hence are ghost-like.

   Although for physical reasons one would probably wish to truncate out the
ghost-like massive modes, there may be circumstances where it could be of
interest to retain them.  It has, for example, been suggested that the
retention of the logarithmic modes in critical gravity could give rise to
an interesting relation to a dual three-dimensional logarithmic CFT on the
AdS$_4$ boundary \cite{Bergshoeff:2011ri}.  A preliminary investigation
of this idea has been initiated in \cite{behamero}, where a toy model
with a scalar field satisfying a fourth-order field equation has been
considered.

   The extensions beyond critical supergravity, i.e. the theories
where the parameter $b$ characterising the Weyl-squared action satisfies
$b\ge 1/a^2$, may provide a family of toy models for renormalisable 
supergravities without ghosts, provided that one truncates out the
negative mass-squared spin-2 modes. It would be interesting to investigate
further the properties of these theories at the quantum level.

\section*{Acknowledgements}

We are grateful to Andy Strominger and Paul Townsend for
helpful discussions. The research of C.N.P.
is supported in part by DOE grant DE-FG03-95ER40917, and the research 
of E.S. and L.W. 
is supported in part by NSF grants PHY-0555575 and PHY-0906222..

\newpage

\appendix

\section{Notation and Conventions}

The $\sigma$-matrices satisfy
\begin{equation}
\sigma^a\bar\sigma^b+\sigma^b\bar\sigma^a=-2\eta^{ab}\,.
\end{equation}
Other useful relations are
\begin{eqnarray}
\varepsilon^{abcd}\sigma_{cd} =-2i\sigma^{ab}\,,\quad
\varepsilon^{abcd}\bar\sigma_{cd}= 2i\bar\sigma^{ab}\,,\quad
\sigma^{abc}=i\varepsilon^{abcd}\sigma_d\,, \quad
\bar\sigma^{abc}=-i\varepsilon^{abcd}\bar\sigma_d\,,
\end{eqnarray}
and
\begin{equation}
\mathrm{tr}(\sigma^{ab}\sigma_{cd})=-4\delta^a_{[c}\delta^b_{d]}-2i\varepsilon^{ab}{}_{cd}\,.
\end{equation}
Dirac gamma-matrices satisfying
\begin{equation}
\gamma^a\gamma^b+\gamma^b\gamma^a=2\eta^{ab}
\end{equation}
are constructed as
\begin{equation}
\gamma^a=\left(\begin{array}{cc}
0&i\sigma^a\\
i\bar\sigma^a&0
\end{array}\right)\,.
\end{equation}
We also have
\begin{equation}
\gamma^5=i\gamma^0\gamma^1\gamma^2\gamma^3=\left(\begin{array}{cc}
-1&0\\
0&1
\end{array}\right)\,.
\end{equation}

\section{$D=4$, ${\cal N}=1$ supergravity}

In this Appendix we will describe the superspace constraints of $D=4$,
${\cal N}=1$ supergravity. To facilitate the comparison to other superspace
literature we will use the notation of Wess and Bagger \cite{Wess:1992cp},
which differs slightly from that used in the rest of the paper. In
particular, letters from the beginning of the alphabet denote tangent space
indices, lower case Latin indices are vector indices, while lower case
Greek indices are spinor indices and capital Latin indices run over
both (in the rest of the paper coordinate vector indices are denoted
$\mu,\nu,\ldots$). The Bianchi identities for the torsion and curvature read
\begin{eqnarray}
D_{[A}T_{BC]}{}^D+T_{[AB}{}^ET_{|E|C]}{}^D+R_{[ABC]}{}^D=0\\
D_{[A}R_{BC]}{}^{DE}+T_{[AB}{}^FR_{|F|C]}{}^{DE}=0\,.
\end{eqnarray}
In the next section we will describe their solution up to mass dimension 3/2, with some results that we will need also at dimension 2 and 5/2. The superspace covariant derivative satisfies
\begin{equation}
[D_A,D_B]=-T_{AB}{}^CD_C+\frac{1}{2}R_{AB}{}^{cd}\ell_{cd}\,,
\end{equation}
where
\begin{equation}
\ell_{cd}V_a=2\eta_{a[c}V_{d]}\qquad \ell_{cd}\psi_\alpha=-\frac{1}{2}(\sigma_{cd})_\alpha{}^\beta\psi_\beta\,,
\end{equation}
on a vector and spinor respectively. This means that the spin-connection satisfies
\begin{equation}
\Omega_\alpha{}^\beta=-\frac{1}{4}\Omega^{cd}(\sigma_{cd})_\alpha{}^\beta
\end{equation}
and similarly for dotted spinor indices.

\subsection{Supergravity constraints}
The non-vanishing components of the torsion and curvature, organized according to mass-dimension, are
\subsubsection*{Dimension 0}
\begin{equation}
T_{\alpha\dot\beta}{}^a=-i\sigma^a_{\alpha\dot\beta}\,.
\end{equation}

\subsubsection*{Dimension 1}
\begin{eqnarray}
T_{a\dot\alpha}{}^\beta=i(\sigma_a)^\beta{}_{\dot\alpha}\RR\,,\qquad
T_{a\alpha}{}^{\dot\beta}=i(\bar\sigma_a)^{\dot\beta}{}_\alpha \RR^\dagger
\end{eqnarray}
and
\begin{eqnarray}
T_{a\alpha}{}^\beta= 2i\delta_\alpha^\beta
G_a-i(\sigma_{ab})_\alpha{}^\beta G^b\,,\qquad
T_{a\dot\alpha}{}^{\dot\beta}=
2i\delta_{\dot\alpha}^{\dot\beta}G_a+i
 (\bar\sigma_{ab})^{\dot\beta}{}_{\dot\alpha} G^b\,,
\end{eqnarray}
where $\RR$ is a scalar superfield and $G_a$ is a real vector
superfield whose lowest components are the auxiliary fields of the
so-called old minimal formulation.
Note in particular the constraint
\begin{equation}
T_{ab}{}^c=0
\end{equation}
which determines the spin-connection.

The curvature components are
\begin{eqnarray}
R_{\alpha\beta cd}=-2(\sigma_{cd})_{\alpha\beta} \RR^\dagger\,,\quad
R_{\dot\alpha\dot\beta cd}=-2(\bar\sigma_{cd})_{\dot\alpha\dot\beta}\RR\,,\quad
R_{\alpha\dot\beta cd}=
-2(\sigma_{bcd})_{\alpha\dot\beta} G^b
\,.
\end{eqnarray}

\subsubsection*{Dimension 3/2}
At this dimension one finds that $\RR$ is chiral,
\begin{equation}
D_{\dot\alpha}\RR=0\,,\qquad D_\alpha \RR^\dagger=0\,,
\end{equation}
as well as
\begin{eqnarray}
\label{eq:DR}
D_\alpha \RR&=&-\frac{1}{6}(\sigma^{cd})_{\alpha\gamma}T_{cd}{}^\gamma
\nn\\
D_{\dot\alpha}\RR^\dagger&=&\frac{1}{6}(\bar\sigma^{cd})_{\dot\alpha\dot\gamma}T_{cd}{}^{\dot\gamma}
\nn\\
D_\alpha G_a&=&\frac{1}{48}(3(\sigma^{cd}\sigma^a)_{\alpha\dot\gamma}T_{cd}{}^{\dot\gamma}-(\sigma^a\bar\sigma^{cd})_{\alpha\dot\gamma}T_{cd}{}^{\dot\gamma})
\nn\\
D_{\dot\alpha}G_a&=&\frac{1}{48}(3(\bar\sigma^{cd}\bar\sigma^a)_{\dot\alpha\gamma}T_{cd}{}^\gamma-(\bar\sigma^a\sigma^{cd})_{\dot\alpha\gamma}T_{cd}{}^\gamma)\,.
\end{eqnarray}
The curvature components of this dimension are
\begin{eqnarray}
R_{\alpha bcd}&=&
\frac{i}{2}(\sigma_b)_{\alpha\dot\gamma}T_{cd}{}^{\dot\gamma}
-\frac{i}{2}(\sigma_d)_{\alpha\dot\gamma}T_{bc}{}^{\dot\gamma}
-\frac{i}{2}(\sigma_c)_{\alpha\dot\gamma}T_{db}{}^{\dot\gamma}
\nn\\
R_{\dot\alpha bcd}&=&
-\frac{i}{2}(\bar\sigma_b)_{\dot\alpha\gamma}T_{cd}{}^\gamma
+\frac{i}{2}(\bar\sigma_d)_{\dot\alpha\gamma}T_{bc}{}^\gamma
+\frac{i}{2}(\bar\sigma_c)_{\dot\alpha\gamma}T_{db}{}^\gamma\,.
\end{eqnarray}

\subsubsection*{Dimension 2}
One finds
\begin{eqnarray}
D_\alpha T_{bc}{}^\gamma
&=&
-2i\delta_\alpha^\gamma G_{bc}
+2i(\sigma_{d[b})_\alpha{}^\gamma D_{c]}G^d
+2(\sigma_{bc})_\alpha{}^\gamma G^2
+4(\sigma_{d[b})_\alpha{}^\gamma G_{c]}G^d
+2(\sigma_{bc})_\alpha{}^\gamma \RR\RR^\dagger
\nonumber\\
&&{}
+\frac{1}{4}R_{bc}{}^{de}(\sigma_{de})_\alpha{}^\gamma
\\
D_{\dot\alpha}T_{bc}{}^\gamma
&=&
2i(\sigma_{[b})^\gamma{}_{\dot\alpha}D_{c]}\RR
+16(\sigma_{[b})^\gamma{}_{\dot\alpha}G_{c]}\RR
+4(\sigma_{bc}\sigma^e)^\gamma{}_{\dot\alpha}G_e\RR\,,
\end{eqnarray}
where $G_{ab}=2D_{[a}G_{b]}$ is the field strength of $G_a$ and similar expressions for $T_{bc}{}^{\dot\gamma}$.

In terms of the superfield
\begin{equation}
W^{\alpha\beta\gamma}=(\sigma^{bc})^{(\alpha\beta}T_{bc}{}^{\gamma)}
\end{equation}
this implies
\begin{eqnarray}
\label{eq:DW}
D_\alpha W^{\beta\gamma\delta}&=&-3i\delta_\alpha^{(\delta}(\sigma^{bc})^{\beta\gamma)}G_{bc}
+\frac{1}{4}R_{bc}{}^{de}(\sigma_{de})_\alpha{}^{(\delta}(\sigma^{bc})^{\beta\gamma)}
\\
D_{\dot\alpha}W^{\alpha\beta\gamma}&=&0\,,
\end{eqnarray}
in particular $W^{\alpha\beta\gamma}$ is a chiral superfield. Similar relations hold for $\bar W^{\dot\alpha\dot\beta\dot\gamma}$.

Using the equation for $D_\alpha \RR$ as well as that for
$D_\alpha T_{bc}{}^\gamma$ one finds
\begin{eqnarray}
\label{eq:D2R}
D^\alpha D_\alpha \RR=
   -\frac{1}{6}(R_{ab}{}^{ab}-12iD_aG^a+24G^2+48\RR\RR^\dagger)\,.
\end{eqnarray}
Similarly one can compute two spinor derivatives on
$\RR^\dagger$ and $G_a$ but we will not do this here as we will not need them.

\subsubsection*{Dimension 5/2}
From the Bianchi identities one finds that
\begin{eqnarray}
D_\alpha R_{bc}{}^{de}
&=&
-2D_{[b}R_{c]\alpha}{}^{de}
-2iG_{[b}R_{c]\alpha}{}^{de}
+2i(\sigma_{[b}\bar\sigma^f)_\alpha{}^\beta R_{c]\beta}{}^{de}G_f
+2i(\sigma_{[b})_\alpha{}^{\dot\beta} R_{c]\dot\beta}{}^{de}\RR^\dagger
\nonumber\\
&&{}
+2(\sigma^{de})_{\alpha\beta}T_{bc}{}^\beta \RR^\dagger
-2(\sigma_{def})_{\alpha\dot\beta}T_{bc}{}^{\dot\beta} G^f\,.
\end{eqnarray}
Using the expression for the dimension 3/2 curvatures this implies that
\begin{eqnarray}
(\sigma^{bc})^{(\beta\gamma}(\sigma_{de})^{\delta)\alpha}D_\alpha R_{bc}{}^{de}
&=&
40W^{\beta\gamma\delta}\RR^\dagger
+10(\sigma^{bc})^{(\beta\gamma}(\sigma^d)^{\delta)}{}_{\dot\beta}
(
iD_bT_{cd}{}^{\dot\beta}
-T_{cd}{}^{\dot\beta}G_b
+2T_{bc}{}^{\dot\beta}G_d
)
\nonumber\\
\end{eqnarray}
Using this expression one computes
\begin{eqnarray}
D^\alpha D_\alpha W^{\beta\gamma\delta}&=&
-10W^{\beta\gamma\delta}\RR^\dagger
+2(\sigma^{bc})^{(\beta\gamma}(\sigma^d)^{\delta)}{}_{\dot\beta}
(
iD_dT_{bc}{}^{\dot\beta}
-4T_{cd}{}^{\dot\beta}G_b
-7T_{bc}{}^{\dot\beta}G_d
)\,.
\end{eqnarray}
One could also derive other relations from the Bianchi identities but we will not need more than these here.

In order to compute the Weyl-squared invariant we need two spinor derivatives of $W^2=W^{\beta\gamma\delta}W_{\beta\gamma\delta}$. With a bit of work one finds
\begin{eqnarray}
D^\alpha D_\alpha(W^2)&=&
-\frac{2}{3}
(
R_{ab}{}^{cd}R^{ab}{}_{cd}
+2R_{ab}{}^{ab}R_{cd}{}^{cd}
+5R_{ab}{}^{cd}R_{cd}{}^{ab}
-12R_{ab}{}^{ac}R_{cd}{}^{bd}
)
+96G^{ab}G_{ab}
\nonumber\\
&&{}
-20W^2\RR^\dagger
+4(\sigma^{ab})^{(\alpha\beta}(\sigma^c)^{\gamma)}{}_{\dot\beta}
(
iD_cT_{ab}{}^{\dot\beta}
-4T_{bc}{}^{\dot\beta}G_a
-7T_{ab}{}^{\dot\beta}G_c
)W_{\alpha\beta\gamma}
\nonumber\\
&&{}
-\frac{i}{3}\varepsilon^{abcd}
(
R^{ef}{}_{ab}R_{efcd}
+R_{ab}{}^{ef}R_{cdef}
+4R_{ab}{}^{ef}R_{efcd}
-8R_{ea}{}^{ef}R_{fbcd}
-144G_{ab}G_{cd}
)\,.
\nonumber\\
\label{eq:D2W2}
\end{eqnarray}
Note that the terms in the last line are imaginary an will therefore not contribute to the action.

\subsection{Components}
Here we collect some component results which we need. The lowest component 
of the superfields $\RR$ and $G_a$ are the auxiliary fields of the old 
minimal formulation of $D=4$ supergravity,
\be
\RR| = \frac{1}{6}\MM\ ,\qquad G_a| = \frac{1}{6}A_a\ .
\ee
(The vector field $A_a$ is customarily called $b_a$ in the superspace 
literature.)
The gravitino is defined as the lowest component of the spinorial supervielbein
\begin{equation}
E_m{}^\alpha|=\psi_m{}^\alpha\,.
\end{equation}
Using this fact, the gravitino field-strength $\psi_{ab}=2D_{[a}\psi_{b]}$ 
can be written
\begin{eqnarray}
\psi_{ab}{}^\gamma
&\equiv&
e_b{}^ne_a{}^mT_{mn}{}^\gamma|
=
T_{ab}{}^\gamma|
-i\psi_{[a}^\gamma\,A_{b]}
+\frac{i}{3}(\sigma_{[a}\sigma^c\psi_{b]})^\gamma\,A_c
-\frac{i}{3}(\sigma_{[a}\bar\psi_{b]})^\gamma\,\MM
\\
\bar\psi_{ab}{}^{\dot\gamma}
&\equiv&
e_b{}^ne_a{}^mT_{mn}{}^{\dot\gamma}|
=
T_{ab}{}^{\dot\gamma}|
+i\bar\psi_{[a}^{\dot\gamma}\,A_{b]}
-\frac{i}{3}(\bar\sigma_{[a}\sigma^c\bar\psi_{b]})^{\dot\gamma}\,A_c
-\frac{i}{3}(\bar\sigma_{[a}\psi_{b]})^{\dot\gamma}\,\bar \MM\,,
\end{eqnarray}
which defines the 'covariantized' gravitino field strength
\begin{eqnarray}
\psi_{ab}{}^{(cov)\gamma}
&\equiv&
T_{ab}{}^\gamma|
=
\psi_{ab}{}^\gamma
+i\psi_{[a}^\gamma\,A_{b]}
-\frac{i}{3}(\sigma_{[a}\bar\sigma^c\psi_{b]})^\gamma\,A_c
+\frac{i}{3}(\sigma_{[a}\bar\psi_{b]})^\gamma\,\MM
\nonumber\\
\bar\psi_{ab}{}^{(cov)\dot\gamma}
&\equiv&
T_{ab}{}^{\dot\gamma}|
=
\bar\psi_{ab}{}^{\dot\gamma}
-i\bar\psi_{[a}^{\dot\gamma}\,A_{b]}
+\frac{i}{3}(\bar\sigma_{[a}\sigma^c\bar\psi_{b]})^{\dot\gamma}\,A_c
+\frac{i}{3}(\bar\sigma_{[a}\psi_{b]})^{\dot\gamma}\,\bar \MM\,.
\label{eq:psicov}
\end{eqnarray}

For the Riemann tensor, $\mathcal R_{ab}{}^{cd}$ which is computed in the
standard way from the spin--connection $\omega^{cd}$,
we find\footnote{The form of $\omega^{cd}$ can be found from the
constraint $T_{ab}{}^c=0$ but we will not need its explicit form. Note
that it will contain $\psi^2$-terms but these will not contribute to
the equations of motion in our case.}
\begin{eqnarray}
\mathcal R_{ab}{}^{cd}&\equiv&e_b{}^ne_a{}^mR_{mn}{}^{cd}|
=
R_{ab}{}^{cd}|
+i\psi_{[a}\sigma_{b]}\bar\psi_{cd}^{(cov)}
-i\psi_{[a}\sigma^d\bar\psi_{b]c}^{(cov)}
+i\psi_{[a}\sigma^c\bar\psi_{b]d}^{(cov)}
\nonumber\\
&&{}
+i\bar\psi_{[a}\bar\sigma_{b]}\psi_{cd}^{(cov)}
-i\bar\psi_{[a}\bar\sigma^d\psi_{b]c}^{(cov)}
+i\bar\psi_{[a}\bar\sigma^c\psi_{b]d}^{(cov)}
-\frac{1}{3}\psi_a\sigma_{cd}\psi_b\,\bar \MM
-\frac{1}{3}\bar\psi_a\bar\sigma_{cd}\bar\psi_b\,\MM
\nonumber\\
&&{}
-\frac{2}{3}\psi_{[a}\sigma^{cde}\bar\psi_{b]}\,A_e\,,
\end{eqnarray}
which gives
\begin{equation}
\label{eq:Rabab}
R_{ab}{}^{ab}|
=
\mathcal R
-2i\psi^a\sigma^b\bar\psi_{ab}^{(cov)}
-2i\bar\psi^a\bar\sigma^b\psi_{ab}^{(cov)}
+\frac{1}{3}\bar\psi_a\bar\sigma^{ab}\bar\psi_b\,\MM
+\frac{1}{3}\psi_a\sigma^{ab}\psi_b\,\bar \MM
+\frac{2}{3}\psi_a\sigma^{abc}\bar\psi_b\,A_c\,.
\end{equation}

\subsection{Supersymmetry transformations}
For completeness we give also the supersymmetry transformations of the component fields. They are given by
\begin{eqnarray}
\delta e_m{}^a&=&-\epsilon^\beta T_{\beta m}{}^a|+\bar\epsilon^{\dot\beta} T_{\dot\beta m}{}^a|\nonumber\\
\delta\psi_m{}^\alpha&=&-D_m\epsilon^\alpha-\epsilon^\beta T_{\beta m}{}^\alpha|+\bar\epsilon^{\dot\beta} T_{\dot\beta m}{}^\alpha|\nonumber\\
\delta\bar\psi_m{}^{\dot\alpha}&=&-D_m\bar\epsilon^{\dot\alpha}-\epsilon^\beta T_{\beta m}{}^{\dot\alpha}|+\bar\epsilon^{\dot\beta} T_{\dot\beta m}{}^{\dot\alpha}|\nonumber\\
\delta \MM&=&-6\epsilon^\alpha D_\alpha \RR|\nonumber\\
\delta\bar \MM&=&6\bar\epsilon^{\dot\alpha}D_{\dot\alpha}\RR^\dagger|\nonumber\\
\delta A_a&=&-6(\epsilon^\alpha D_\alpha-\bar\epsilon^{\dot\alpha}D_{\dot\alpha})G_a|\,.
\end{eqnarray}
Using the superspace constraints in section B.1 and the component results in section B.2 we find
\begin{eqnarray}
\delta e_m{}^a&=&-i\bar\psi_m\bar\sigma^a\epsilon-i\psi_m\sigma^a\bar\epsilon\nonumber\\
\delta\psi_m{}^\alpha&=&-D_m\epsilon^\alpha
+\frac{i}{3}\epsilon^\alpha\,A_m+\frac{i}{6}(\sigma_{mb}\, 
\epsilon)_\alpha\,A^b
-\frac{i}{6}(\sigma_m\, \bar\epsilon)^\alpha\,\MM
\nonumber\\
\delta\bar\psi_m{}^{\dot\alpha}&=&-D_m\bar\epsilon^{\dot\alpha}
-\frac{i}{3}\bar\epsilon^{\dot\alpha}\,A_m-\frac{i}{6}(\bar\sigma_{mb}\,
\bar\epsilon)^{\dot\alpha}\,A^b
-\frac{i}{6}(\bar\sigma_m\, \epsilon)^{\dot\alpha}\,\bar \MM
\nonumber\\
\delta \MM&=&-\epsilon\, \sigma^{cd}\psi_{cd}^{(cov)}
\nonumber\\
\delta\bar \MM&=&-\bar\epsilon\, \bar\sigma^{cd}\bar\psi_{cd}^{(cov)}
\nonumber\\
\delta A_a&=&
\frac{1}{8}(3\bar\epsilon\, 
\bar\sigma^{cd}\bar\sigma^a\psi_{cd}^{(cov)}-\bar\epsilon\,
\bar\sigma^a\sigma^{cd}\psi_{cd}^{(cov)}
-3\epsilon\, 
\sigma^{cd}\sigma^a\bar\psi_{cd}^{(cov)}+\epsilon\,
\sigma^a\bar\sigma^{cd}\bar\psi_{cd}^{(cov)})\,.
\end{eqnarray}

\subsection{Quadratic gravitino terms from Weyl$^2$ invariant in AdS$_4$}
Supersymmetric Lagrangians can be constructed as
\begin{equation}
e^{-1}\mathcal L=\left(\ft12 D^\alpha D_\alpha+i(\bar\psi_a\sigma^a)^\alpha
D_\alpha+\bar \MM
  +\bar\psi_a\bar\sigma^{ab}\bar\psi_b\right)r|+\mathrm{h.c.}\,,
\end{equation}
where $r$ is a chiral superfield. Taking $r=-\ft14W^{\alpha\beta\gamma}W_{\alpha\beta\gamma}$ gives the Weyl-squared invariant. In this section we shall 
compute the terms quadratic in the gravitino in the AdS$_4$ background 
given by $\MM=3a$ and $\mathcal R_{ab}{}^{cd}$ given in (\ref{AdS}).

Using these expressions it is not hard to see that the curvature terms in (\ref{eq:D2W2}) do not give any contribution to the quadratic gravitino terms in the action in this background. Similarly the term $G_{ab}G^{ab}|$ can not give any quadratic gravitino contribution. Using (\ref{eq:D2W2}) and (\ref{eq:DW}) we find that the quadratic gravitino terms in the AdS$_4$ background are
\begin{eqnarray}
e^{-1}\mathcal L_\psi&=&\frac{a}{2}W^2|
-\frac{i}{2}(\sigma^{ab})^{(\alpha\beta}(\sigma^c)^{\gamma)}{}_{\dot\beta}D_cT_{ab}{}^{\dot\beta}|\,W_{\alpha\beta\gamma}|
+\mathrm{h.c.}\,.
\end{eqnarray}
Using the fact that
\begin{equation}
D_aT_{bc}{}^{\dot\gamma}|=D_a\bar\psi_{bc}{}^{\dot\gamma}
  +\frac{i}{3}(\bar\sigma_{[b}D_{|a|}\psi_{c]})^{\dot\gamma}\bar \MM+\ldots\,,
\end{equation}
together with
\begin{equation}
W^{\alpha\beta\gamma}|=(\sigma^{de})^{(\alpha\beta}\psi_{de}^{\gamma)}+\ldots\,,
\end{equation}
where $\ldots$ denotes terms that vanish in the AdS$_4$ background when expanded to linear order the two $W^2$-terms cancel and we find
\begin{eqnarray}
e^{-1}\mathcal L_\psi&=&
\frac{i}{2}(\sigma^{ab})^{\alpha\beta}(D_c\bar\psi_{ab}\bar\sigma^c)^\gamma\,(\sigma^{de})_{(\alpha\beta}\psi_{de\gamma)}
+\mathrm{h.c.}\,.
\end{eqnarray}
Simplifying and dropping total derivatives we finally arrive at the Lagrangian
\begin{eqnarray}
e^{-1}\mathcal L_\psi&=&
\frac{4}{3}\left(
iD_d\bar\psi^{ab}\bar\sigma^d\psi_{ab}
-i\bar\psi^{ab}\bar\sigma^dD_d\psi_{ab}
+iD_d\bar\psi_{a}{}^c\bar\sigma^{abd}\psi_{bc}
+i\bar\psi_{b}{}^c\bar\sigma^{abd}D_d\psi_{ac}
\right)\,.
\end{eqnarray}

\section{Relations between regular modes in AdS$_4$}

AdS$_4$ admits four Killing spinors $\ep_+$ and four Killing spinors $\ep_-$,
satisfying
\be
\nabla_\mu\ep_+ = \ft12 a\, \gamma_\mu\, \ep_+\,,\qquad
\nabla_\mu\ep_- = -\ft12 a\, \gamma_\mu\, \ep_-\,.\label{kspinors}
\ee
These can be used in order to map between modes of different spins.  We
begin by defining the second-order operators, and eigenvalues, for each spin:
\bea
\hbox{Spin 0}:&& \Delta_0\, \phi \equiv -\square\phi = \lambda_0\, \phi\,,\nn\\
\hbox{Spin\ }\ft12:&& \Dslash\psi\equiv \gamma^\mu \nabla_\mu \psi =
     \lambda_{1/2}\, \psi\,,\nn\\
\hbox{Spin 1}:&& \Delta_1 V_\mu \equiv -\square V_\mu + R_{\mu\nu}\, V^\nu =
   \lambda_1\, V_\mu\,,\nn\\
\hbox{Spin\ }\ft32:&& \Dslash\psi_\mu\equiv \gamma^\nu \nabla_\nu \psi_\mu =
     \lambda_{3/2}\, \psi_\mu\,,\nn\\
\hbox{Spin 2}:&& \Delta_L h_{\mu\nu}\equiv -\square h_{\mu\nu} -
  2 R_{\mu\rho\nu\sigma}\, h^{\rho\sigma} + R_{\mu\rho}\, h^\rho{}_\nu
  + R_{\nu\rho}\, h_\mu{}^\rho = \lambda_L\, h_{\mu\nu}\,.
\eea
Note that we assume transverse and traceless conditions for the modes
of spins 1, $\ft32$ and 2, and so
\be
\nabla^\mu V_\mu=0,\qquad \nabla^\mu\psi_\mu=0\,,\qquad
\gamma^\mu\psi_\mu=0\,,\qquad \nabla^\mu h_{\mu\nu}=0\,,\qquad
h^\mu{}_\mu=0\,.\label{ttcon}
\ee
In the AdS$_4$ background, and setting $a=1$ for convenience, we have
\be
R_{\mu\nu\rho\sigma}= -  g_{\mu\rho}\, g_{\nu\sigma} +
      g_{\mu\sigma}\, g_{\nu\rho}\,,\qquad
R_{\mu\nu}= -3  g_{\mu\nu}\,,
\ee
and so the spin 1 and spin 2 operators become
\be
\Delta_1 = -\square -3 \,,\qquad
\Delta_L =-\square -8 \,.
\ee

By default, we shall consider the case where the Killing spinors $\ep_+$
are used for relating the various modes, and for brevity we shall just
denote these by $\ep$.  We find that the relations between the modes
are implemented as follows:
\bea
\psi&=& \phi\, \ep + \fft1{\lambda_{1/2}+1}\, \nabla_\mu \phi\, \gamma^\mu\ep
\,,\nn\\
V_\mu &=& \bar\ep \gamma_\mu\psi -
  \fft1{\lambda_{1/2} +\fft32 }\, \bar\ep\,\nabla_\mu\,\psi\,,\nn\\
\psi_\mu&=& V_\mu\,\ep + \ft14  c (1-2\lambda_{3/2}-2\lambda_{3/2}^2)\,
\gamma_{\mu\nu}\, V^\nu\,\ep + c (1+\lambda_{3/2})\,
\nabla_\nu V_\mu\, \gamma^\nu\,\ep -c\, \nabla_\mu V_\nu\, \gamma^\nu\, \ep
\nn\\
&&-\fft1{2\lambda_{3/2}}\, \gamma_{\mu\nu\rho}\, \nabla^\nu V^\rho\, \ep +
\ft12 c\, \nabla_{(\mu}\nabla_{\nu)} V_\rho \, \gamma^{\nu\rho}\, \ep\,,\nn\\
h_{\mu\nu} &=& \bar\ep \gamma_{(\mu}\, \psi_{\nu)} -\fft{2}{2\lambda_{3/2}+5}\,
\bar\ep \nabla_{(\mu}\, \psi_{\nu)}\ ,
\eea
where $c^{-1}=\lambda_{3/2}\, (2+\lambda_{3/2})$.  (The relative coefficients
between the terms in each expression are uniquely determined by requiring
that the irreducibility conditions in (\ref{ttcon}) hold, and that the
constructions should map eigenfunctions into eigenfunctions.) These formulae 
furnish a systematic way of constructing the spin $1/2$, 1, $3/2$ and 2 
solutions, starting from the spin $0$ solution, and the knowledge of 
the Killing spinor. The spin $0$ solution has been studied in great detail 
in \cite{Starinets:1998dt}. Alternatively, starting from the spin $2$ 
solution, we can obtain from it the spin $3/2$, 1, $1/2$ and 0 
solutions by employing the formulae

\bea
\psi_\mu &=& h_{\mu\nu}\, \gamma^\nu\, \ep -\fft1{\lambda_{3/2}}\,
\nabla_\rho h_{\mu\nu}\, \gamma^{\nu\rho}\, \ep\,,\nn\\
V_\mu &=& \bar\ep \psi_\mu\,,\nn\\
\psi &=& V_\mu\,\gamma^\mu\,\ep +
\fft1{\lambda_{1/2}}\, \nabla_\mu V_\nu\, \gamma^{\mu\nu}\, \ep\,, \nn\\
\phi &=& \bar\ep \psi\ .
\eea

The corresponding relations between the eigenvalues are
\bea
\lambda_0 &=& -\lambda_{1/2}^2 + \lambda_{1/2} +2\,,\nn\\
\lambda_1 &=& -\lambda_{1/2}^2 -\lambda_{1/2}\,,\nn\\
\lambda_1 &=& -\lambda_{3/2}^2 + \lambda_{3/2}\,,\nn\\
\lambda_L &=& -\lambda_{3/2}^2 - \lambda_{3/2}-4\,.
\eea
(If $\ep_-$ is used instead of $\ep_+$, the effect is to reverse the signs
of the fermion eigenvalues in these expressions.)



\newpage

\begin{thebibliography}{99}

\bibitem{stelle1} K.S. Stelle,
{\it Renormalization of higher derivative quantum gravity},
Phys. Rev. {\bf D16}, 953 (1977).

\bibitem{stelle2} K.S. Stelle,
{\it Classical gravity with higher derivatives},
Gen. Rel. Grav. {\bf 9}, 353 (1978).


\bibitem{lupo} H. L\"u and C.N. Pope,
{\it Critical gravity in four dimensions},
Phys. Rev. Lett. {\bf 106}, 181302 (2011),
arXiv:1101.1971 [hep-th].

\bibitem{porrob} M. Porrati and M.M. Roberts,
{\it Ghosts of critical gravity},
arXiv:1104.0674.

\bibitem{lilulu} H. Liu, H. L\"u and M. Luo,
{\it On black hole stability in critical gravities},
arXiv:1104.2623.

\bibitem{maldaconf}
  J. Maldacena,
{\it Einstein gravity from conformal gravity,}
arXiv:1105.5632 [hep-th].

\bibitem{lupapo} H. L\"u, Y. Pang and C.N. Pope,
{\it Conformal gravity and extensions of critical gravity},
arXiv:1106.4657 [hep-th].

\bibitem{Stelle:1978ye}
K.S. Stelle and P.C. West,
{\it Minimal auxiliary fields for supergravity},
Phys. Lett. {\bf B74} (1978) 330.

\bibitem{Ferrara:1978em}
S. Ferrara and P. van Nieuwenhuizen,
{\it The auxiliary fields of supergravity},
Phys. Lett. {\bf B74} (1978) 333.

\bibitem{Ferrara:1988pd}
S. Ferrara, S. Sabharwal and M. Villasante,
{\it Curvatures and Gauss-Bonnet theorem in new minimal supergravity},
Phys. Lett. {\bf B205} (1988) 302.

\bibitem{Cecotti:1987qe}
S. Cecotti, S. Ferrara, M. Porrati and S. Sabharwal,
{\it New minimal higher derivative supergravity coupled to matter},
Nucl. Phys. {\bf B306} (1988) 160.

\bibitem{deRoo:1990zm}
M. de Roo, A. Wiedemann and E. Zijlstra,
{\it The construction of $R^2$ actions in $D = 4$, $N=1$ supergravity},
Class. Quant. Grav. {\bf 7} (1990) 1181.

\bibitem{LeDu:1997us}
R. Le Du,
{\it Higher derivative supergravity in $U(1)$ superspace},
Eur. Phys. J.  {\bf C5}, 181 (1998), hep-th/9706058.

\bibitem{Binetruy:2000zx}
P. Binetruy, G. Girardi and R. Grimm,
{\it Supergravity couplings: A geometric formulation},
Phys. Rept. {\bf 343} (2001) 255, hep-th/0005225.

\bibitem{heid} W. Heidenreich,
{\it All linear unitary irreducible representations of de sitter supersymmetry
with positive energy,}
Phys. Lett. {\bf B110} (1982) 461.

\bibitem{breifree} P. Breitenlohner and D.Z. Freedman, {\it Positive energy
in anti-de Sitter backgrounds and gauged extended supergravity},
Phys. Lett. {\bf B139} (1984) 154.


\bibitem{Bergshoeff:2011ri}
E.A. Bergshoeff, O. Hohm, J. Rosseel and P.K. Townsend,
{\it Modes of log gravity},
Phys. Rev. {\bf D83} (2011) 104038, arXiv:1102.4091 [hep-th].

\bibitem{Wess:1992cp}
J. Wess and J. Bagger,
{\it Supersymmetry and supergravity},
Princeton, USA: Univ. Pr. (1992).

\bibitem{Starinets:1998dt}
A. Starinets,
{\it Singleton field theory and Flato-Fronsdal dipole equation},
Lett. Math. Phys. {\bf 50} (1999) 283, math-ph/9809014.

\bibitem{behamero}
E.A. Bergshoeff, S. de Haan, W. Merbis and J. Rosseel,
{\it A non-relativistic logarithmic conformal field theory from a holographic
point of view},
arXiv:1106.6277 [hep-th].


\end{thebibliography}
\end{document}